# Sounding Canvas: Embedding Algorithms in Networked, Sensorial Sound Art

## Authors:

Luciano Ciamarone*, Dora Motèque*, Marco Giordano**

* independent researcher; email: valmovigo@musician.org; ORCID: https://orcid.org/0009-0001-8405-0894
** Dpt. of Information Engineering, Computer Science and Mathematics (DISIM), University of L'Aquila, L'Aquila, Italy; email: marco.giordano3@graduate.univaq.it; ORCID: https://orcid.org/0009-0001-1649-6085

## Abstract

*Sounding Canvas* turns painting into a touch-responsive multimodal installation by embedding capacitive sensors, real-time decision models and networking inside the canvas. Touches trigger spatialised sounds that seem to emanate from the painting itself. The work embeds algorithms in three senses: (1) physically, as sensing and computation concealed behind the artwork; (2) perceptually, via an off-line visual-sonic mapping that aligns a painting's features with sound descriptors; and (3) performatively, through on-line models that shape the live dialogue with visitors and with remote canvases over the network. We describe the artistic rationale and implementation: a CNN-based off-line mapping that curates the sound vocabulary, and two on-line event managers, a high-order Markov model and an LSTM policy, that balance responsiveness with guided exploration. We discuss how these layers make the algorithm perceivable as behaviour rather than code, and how networking turns solitary touch into distributed co-authoring. We conclude with reflections on authorship, agency and evaluation methods for embedded algorithmic artworks. A GitHub repository of the project is available for public access.

## 1. Introduction

Sounding Canvas transforms paintings made with conventional techniques into interactive sonic environments. Capacitive sensors, microcontrollers, and algorithmic processes embedded within the canvas turn visual art into a living sound installation: visitors activate the work through touch, triggering sonic responses that evolve in real time. Each canvas connects to a distributed network, enabling geographically distant works to exchange interaction data and collectively generate a shared soundscape.

The project sits at the intersection of embedded systems, machine learning, and sound art. In its earliest presentation, Sounding Canvas was framed within the context of Internet of Sounds research, focusing on sensor arrays, Raspberry Pi processing, and network protocols. The present article shifts emphasis toward the conceptual and artistic dimensions of the work, aligning it with the *Organised Sound* call for papers on embedding algorithms in music and sound art. Here, *embedding* is understood in multiple senses: the physical embedding of sensors into the canvas; the embedding of visual features into sonic descriptors via convolutional neural networks; and the embedding of probabilistic models, such as Markov chains or Recurrent Neural Networks, into the temporal flow of human–machine interaction.

These layers show how algorithms are not merely hidden engines of computation but central to the aesthetic, performative, and social qualities of the work. Sounding Canvas demonstrates how embedding algorithms within an artistic system reshapes the relationship between artwork, performer, and audience, creating new modes of distributed authorship and networked participation. The following sections position the project within sound art traditions, describe its algorithmic frameworks, and discuss how these embeddings manifest as both technical practice and artistic statement.

The project GitHub repository, which also contains links to video materials, is available at: https://github.com/luciamarock/SoundingCanvas

# 2. Conceptual and Artistic Background

The Sounding Canvas draws upon a lineage of interactive systems that merge the visual and the sonic through embodied, often tactile, interaction.
A number of works have explored the canvas or mural as a responsive surface. Zhang's Conduct San Jose (2019) employed Bare Conductive paint to transform a public mural into a touch-sensitive musical instrument, while Soundwall (2015) combined flat-panel painting-surfaces with Wi-Fi–enabled loudspeakers to create hybrid objects of art and sound. Daniela Voto's Multisensory Interactive Installation (2025) linked Kandinsky-inspired colors and shapes to musical chords and rhythms, offering a synesthetic translation of painting into music. Earlier, Reuter's SoundPaint (2005) provided a systematic framework for color-to-sound mappings, and Doury et al.'s Paint With Music (Google Magenta, 2021) extended this tradition into machine learning, using DDSP to render user drawings as real-time instrumental music.

Beyond painted surfaces, touch-responsive textile and tapestry installations represent a closely related tradition. Briot, Honnet and Strohmeier's *Stymphalian Birds* (DIS 2020) explored the sonification of touch through conductive textile materials, demonstrating how tactile interaction with fabric surfaces can generate sonic responses embedded in the work's physical matter. Buechley's *Tinkering Tinkerer* extended computational craft practices to wall-mounted painted surfaces, foregrounding audience touch as a generative act. More recently, Santos et al. (IDC 2025) and the *TapeStory* project (C&C 2024) applied capacitive weaving techniques to create wall-hung narrative installations in which visitor touch triggers sound and storytelling, establishing interactive textiles as an active medium for the embodied, surface-based sonic experience that the Sounding Canvas also pursues.

The Sounding Canvas also resonates with broader practices of distributed and networked interaction. Lozano-Hemmer's Remote Pulse (2019) exemplifies how embodied gestures can be shared across distance, producing a form of remote presence akin to what our canvases achieve through touch events. Koblin and Echelman's Unnumbered Sparks (2014) and Murray-Browne's Cave of Sounds (2018) demonstrate collaborative frameworks for distributed sound-making, highlighting the social dimension of networked art.

On a technical level, capacitive sensing has been an important foundation for touch-based artworks. Fernandez and Larson's Tune Field (NIME 2021) documents a tangible interface using capacitive sensing and OSC transmission, while Honigman et al. (NIME 2014) provide open-source swept-frequency sensing techniques suitable for interactive paintings. Blazey's Kalimbo (NIME 2017) integrates capacitive touch with gestural sensors, exemplifying multi-modal input strategies. Dannenberg (2021) outlines the O2 protocol for distributed real-time music systems, providing essential guidance for synchronization and WebSocket-based communication as employed in the Sounding Canvas.

Theoretical frameworks further situate the work within cross-modal and aesthetic research. Fink et al. (2024) examine audiovisual congruence in contemporary music and art, challenging assumptions about intuitive mapping and emphasizing embodied perception. The Leonardo Music Journal (2013) and Organised Sound (2009) special issues on sound art articulated site-specificity, multichannel practice, and object-based installations as key categories for situating sound within visual environments. Algorithmic agency frames the Sounding Canvas within current debates on machine creativity: Rutz (2016) describes how algorithms operate as co-creative agents, a perspective reinforced by the ALMAT symposium (2020). Work in computational creativity (Carnovalini & Rodà 2020; Bustos & Pinto 2022) and interactive machine learning (Fiebrink & Cook 2010; Visi & Tanaka 2020) provide methodological precedents for adaptive mappings and reinforcement-driven musical systems, while Neef (2024) examines shifting audience perceptions of AI-generated versus human-authored art, raising tensions around authorship and aesthetic value directly relevant to the Sounding Canvas.

Capacitive sensing transforms each gesture into spatialised and timbrally nuanced audio output: the tactile layer is not merely a trigger for static events but a medium through which users sense and influence emergent sonic behavior. Repeated gestures are interpreted in context, allowing the canvas to carry memory of past interactions and adapt its responses accordingly. LSTM-based adaptive mapping distinguishes short-term from long-term interaction patterns, allowing each canvas to evolve a distinctive sonic personality guided by curated aesthetic constraints. The probabilistic model ensures output is not random but a complex, deterministic function of the system's memory and incoming gesture dynamics. When canvases are networked, gestures on one canvas influence sound generation on another, creating distributed tactile agency in which user input is interpreted and amplified, but never directly replicated, fostering emergent behaviors and collaborative awareness.

The genesis of our audiovisual approach lies in Dora Motèque's semiographic research. Her first impulse emerged from her encounters with Michelangelo Lupone's interactive sculptures at CRM - Centro Ricerche Musicali (Centre for Musical Research) in Rome (Lupone, 2006), where metallic forms respond to touch producing sound and actuating it through the structures of the installation itself. While she admired this transgression of the museum taboo "do not touch", she sought to go further, to create something that would appear precious and fragile, an object that viewers might instinctively hesitate to touch but which would demand touch in order to come fully alive. Painting offered that paradox: the untouchable medium made touch essential.

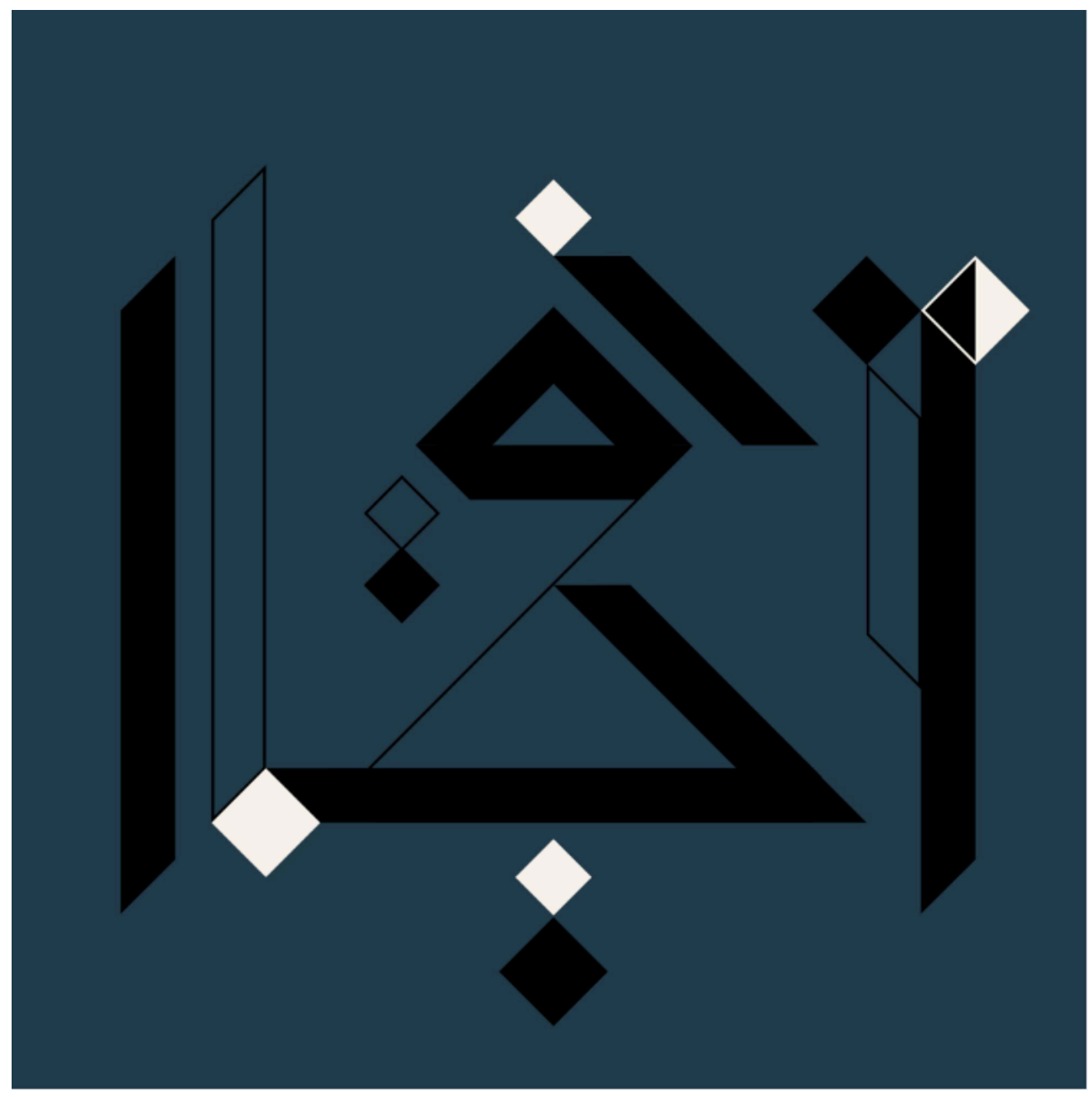

*Fig. 1: Sounding Canvas: "Calligraph of Absence"*

The visual language inscribed onto the canvases is not decorative but procedural (Figure 1). Drawing on her Persian heritage and lifelong practice of calligraphy (خط), together with her training in scenography, Motèque fused the flowing signs of Persian script with the abstract logic of musical notation. In calligraphy, the gestures of line and point already carry an affinity with musical phrasing; here, that affinity becomes material, guiding the hand as if following a score. The resulting semiographic syntax invites tactile exploration while encoding cultural memory and personal biography into symbolic form.

The sonic identity of each Sounding Canvas is rooted in the composer's (Luciano Ciamarone) expertise in translating abstract audio descriptors into expressive musical structures. A custom machine-learning system maps the visual data of the canvases into a specific set of audio descriptors (detailed in Section 5), which serve as the primary creative constraints for the soundscapes. The system is trained on the composer's personal aesthetic preferences rather than a generalized dataset, using a curated selection of 50 pairs of video stills and approximately 20-second audio segments—ranging from ambient textures to foley and archival music sourced from art documentaries—to learn subjective "favorite" associations and produce a symbolic "score" of target descriptors. The composer then interprets these quantities and instills them into the final audio samples. This mapping establishes strong audiovisual congruence: visual gestures, shapes, flows, and densities directly inform temporal, timbral, and spatial aspects of the sound, prioritizing aesthetic coherence and preserving the composer's stylistic signature across modalities.

Early explorations were grounded in the idea of creating an agent capable of learning from human interaction, with an initial framework based on Markov Models chosen for their capacity to formalize evolving interaction patterns. This resonates with earlier traditions in algorithmic composition—Xenakis's stochastic works (Xenakis 1959, 1971) and Boulez's serial explorations (1960, 1976)—in contrast to the indeterminate strategies of John Cage (1957, 1961). Higher-order Markov Models (HOMM) (Raftery 1985; Ching 2004; Salnikov 2016) enabled the system to associate shifting contexts with evolving sonic probabilities, echoing also genetic programming and evolutionary approaches in music (Bentley 1999; Todd 1999; Biles 1994). However, in single-user contexts, Markov chains tended to reinforce repetitive habits, raising the question of how an agent might sustain personality without collapsing into rigidity.
To address this, the system evolved toward Recurrent Neural Networks (RNNs) (Elman 1990), specifically Long Short-Term Memory (LSTM) (Hochreiter 1997) models, which were better suited for distinguishing between short-term and long-term memory. In biological and cognitive models of memory, short-term traces are often erased, while long-term patterns consolidate as “ways of doing” or signatures of personality (Atkinson 1968, Baddeley 1974). Applied to the Sounding Canvas, this meant that not all individual gestures were retained, but rather that persistent trends shaped the evolving behavior of the agent. The result was an emergent authorship, where the system developed a recognizable character across time, informed but not determined by any single interaction.
Comparable strategies can be found in Hanns Holger Rutz’s (2017) investigations of algorithmic agency (e.g., Sound Processes), where systems actively shape the trajectory of interaction, or in Rafael Lozano-Hemmer’s works (2009) where participant input is transformed into algorithmic goals that exceed individual authorship.
These developments reopen long-standing debates on agency and determinism. If an agent, whether algorithmic or biological, acts on the basis of probabilistic rules, memory traces, and reinforcement processes, then the question arises whether it chooses at all, or merely executes an implicit score. This perspective resonates with biological determinism, where human decision-making is framed as the outcome of biochemical and neurological processes (Crick 1994, Libet 1983). In this sense, the Sounding Canvas positions itself not merely as an interactive artwork, but as a microcosm of broader questions concerning agency, memory, and determinism. While we do not presume to offer definitive answers to these fundamental issues, we suggest that meaningful insight, if not resolution, can emerge through the study of interaction itself. It is precisely for this reason that we introduced a layer of connectivity between individual entities.
When networking two or more Sounding Canvases, the LSTM networks function not merely as responsive interfaces but as mediating infrastructure that enables distant users to mutually affect each other's aesthetic experiences through algorithmic interpretation and sound selection.
What begins as a human-machine dialogue evolves into human-machine-machine-human encounters when canvases connect across the network. The algorithm becomes social glue by translating one user's gestural intentions into sonic interventions within another user's performance space. For instance, a gentle touch in Rome might trigger complementary harmonic selections on a Barcelona canvas, requiring users to develop sensitivity toward invisible collaborators whose actions directly influence their local sonic environment.
This cross-affectation through algorithmic mediation creates social responsibility within the aesthetic encounter. Users must attune themselves to remote participants, learning to listen not only to their own canvas but to the algorithmic interpretations of distant gestures manifesting as unexpected sounds within their performance. The technology compels users to develop collaborative awareness, transforming individual expression into collective negotiation. Following Joseph Beuys' (1971, 1973) conception of art as social sculpture that shapes human relationships and consciousness, the Sounding Canvas network forces humans to listen to one another across geographical boundaries. The algorithmic embedding creates conditions where individual aesthetic agency becomes inseparable from collective responsibility, users must consider how their gestures affect distant participants, fostering empathetic engagement with unknown collaborators mediated through machine intelligence that translates intention across space and time.

## 3. System Overview

The Sounding Canvas hardware (Figure 2) integrates capacitive sensing with embedded audio playback to create a responsive interactive artwork. Capacitive sensors, made from copper foil pads of approximately 40 cm², are mounted on the back of the canvas (Figure 3).

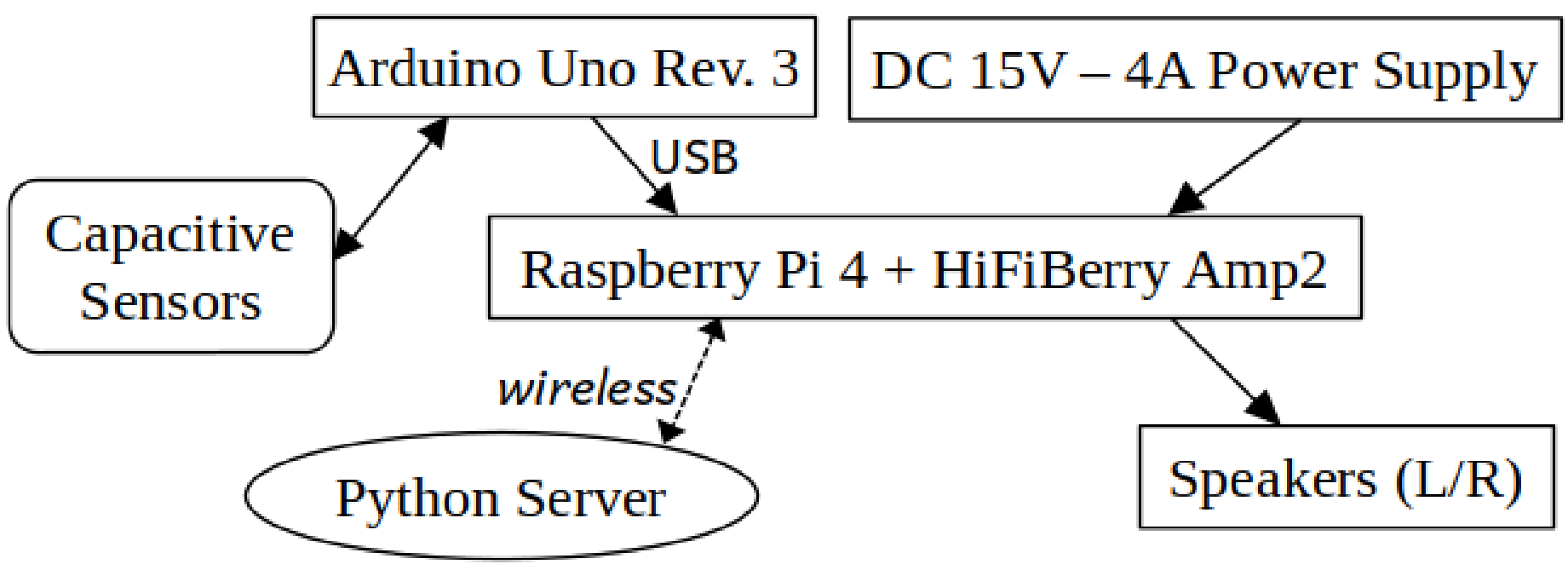

*Fig. 2: High Level block architecture of the system*

Unlike uniform sheets, each pad is cut to match the geometric shapes present in the painting and can be applied in a “wavy” or discontinuous pattern, sometimes with zebra-like interruptions. This deliberate disruption of continuity allows the system to capture fluctuations in the capacitance signal, providing information not only about touch but also about the speed and dynamics of the user’s gestures, which serve as an input to the RNN-based interaction model. Each sensor forms an RC circuit with a 1.4 MΩ resistor, enabling the Arduino Uno Rev3 to reliably detect changes in capacitance. The boards and electronics are mounted on a wooden back panel made of MDF (Figure 4), structured as a sandwich with an internal aluminum foil layer that serves as electromagnetic shielding; this shield is connected to the Arduino’s GND pin to reduce interference. Interaction events are transmitted via USB to a Raspberry Pi 4 Model B with a HiFiBerry Amp2 HAT, which triggers pre-recorded sound samples rather than synthesizing audio. Two 4-inch loudspeakers (20-40 W, 4 Ω) are mounted alongside the electronics, with cabling carefully routed and all components secured with insulating layers and supports. This configuration ensures accurate detection, facilitates maintenance, and allows the system to interpret both spatial and dynamic aspects of user interactions, additionally this specific hardware stack was chosen to decouple the sensing logic from audio playback for better system maintenance, with the Raspberry Pi/HiFiBerry combo providing a practical all-in-one power and amplification solution sufficient for non-instrumental, sample-based interaction.

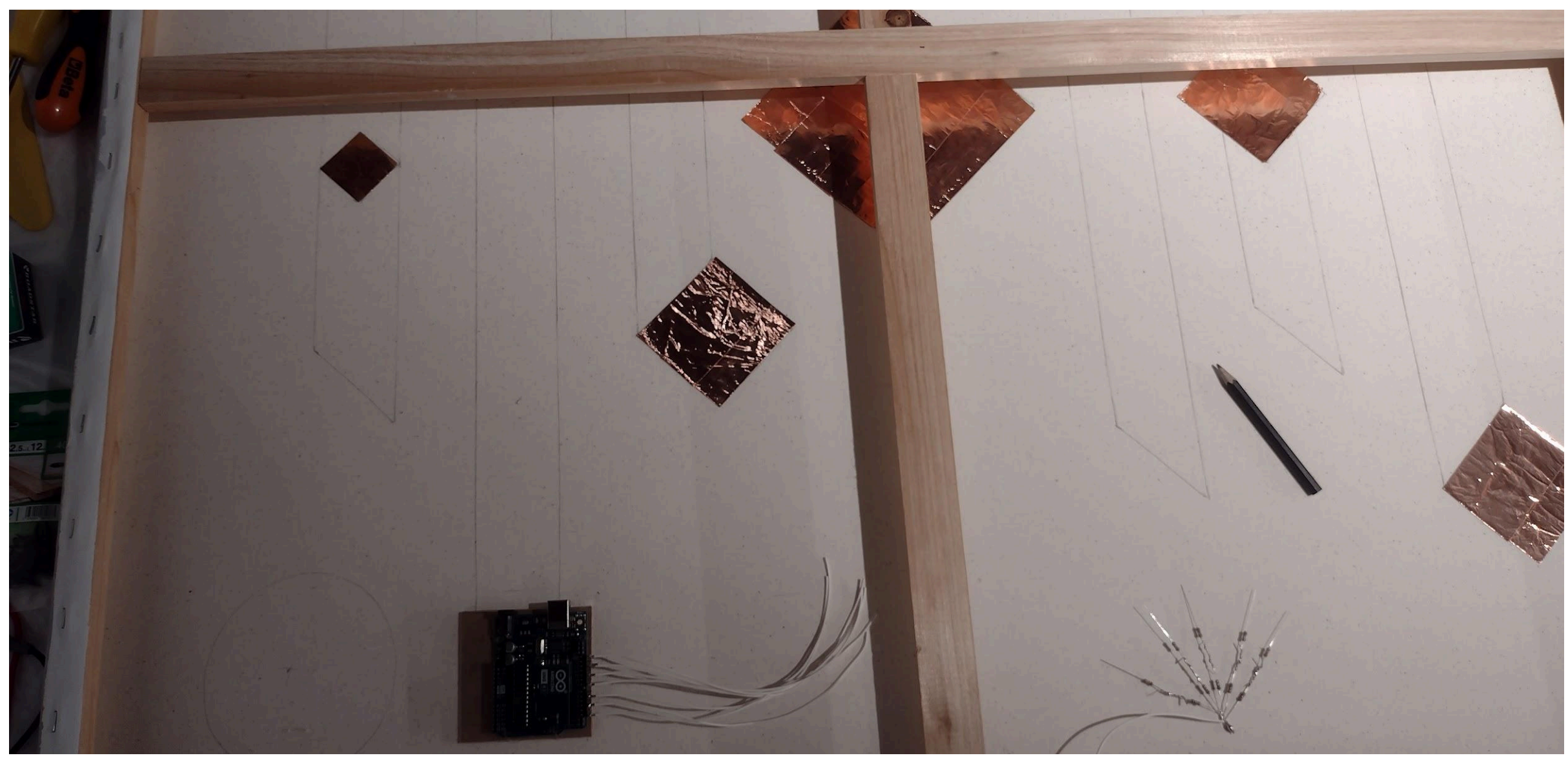

*Fig. 3: Sensor Layer of Sounding Canvas "Silent Score"*

The software architecture of the Sounding Canvas is organized to seamlessly bridge user interaction, real-time sound manipulation, and networked communication, while maintaining artistic coherence. In the codebase (Ciamarone 2024), a configuration file provides a single, user-accessible source for defining all relevant parameters, from Arduino port and touch sensor thresholds to WebSocket server settings and event manager configurations, ensuring flexibility across different installations. The central interface, `sc_helper.py`, abstracts hardware and system details, offering `run.py` (the main entry point) a unified API for reading sensor input, managing thresholds, smoothing signals, and routing events to the appropriate event manager. In the original HOMM-based setup, the `event_manager.py` used a high-order Markov model to generate temporally coherent audio responses by learning transition probabilities from sequences of user interactions; however, it was prone to self-reinforcement in single-user sessions. To overcome this limitation, a Recurrent Neural Network (LSTM) was introduced. The LSTM model, trained on sequences of past sound events, inter-touch intervals, and gesture features, predicts the probability distribution of the next channel the user might activate, allowing the system to dynamically select sounds that maintain engagement and variability. Within the live system, the `MachineAnswer` class encapsulates this logic, maintaining a history of user interactions, calculating goal channels, and evaluating candidate sounds through the LSTM, while respecting timing and gesture features. The entire pipeline, from sensor detection, via `sc_helper.py`, through event management and the LSTM, to sound output, is designed to create each canvas as an autonomous yet responsive musical entity, whose voice and temporal behavior reflect both the user's gestures and the intrinsic "personality" of the installation. Asynchronous coroutines in `run.py` handle local sensor reading and WebSocket communication concurrently, ensuring smooth responsiveness whether the canvas is used solo or in a networked multi-canvas setting, fully preserving the artistic experience.

# 4. Embedding and Performance

The physical act of touching or approaching the canvas creates a subtle, measurable disruption in the embedded capacitive fields. Copper foil sensors, discreetly placed behind the artwork, detect these minute changes in electrical capacitance. The Arduino's CapacitiveSensor library then transforms this analog physical gesture into a continuous stream of raw numerical data. This raw data is not inherently meaningful; rather, the algorithm performs the crucial work of interpretation. It analyzes the oscillating behavior of the sensor stream to draw conclusions, for example, that a user is moving a hand over the painted shapes of canvas rather than keeping it still. When in the case of RNN based events management, this intelligent interpretation is fed into the LSTM network.
The LSTM architecture functions as a predictive temporal engine, utilizing its memory of a user's previous gestural trajectories to forecast subsequent interactions. Rather than maintaining a purely reactive posture, the system uses these predictions to resolve a specific, designer-defined teleology: the pursuit of equilibrium across the tactile surface. In this implementation, "equal wear" is selected as the primary machine objective, serving as a catalyst for a non-hierarchical exploration of the canvas. By anticipating the user's next move, the agent can strategically trigger "attractor" sonic events in underutilized sensor zones, subtly shifting the user's focus toward untouched areas. This creates a state of negotiated agency, where the sonic development is not merely a result of human whim, but a byproduct of the system's internal drive for physical and acoustic symmetry. While other machine goals could be programmed, such as prioritizing specific spectral densities or rhythmic densities, the choice to balance sensor activation was selected both for practical and evaluative reasons. Distributing interaction evenly helps prevent the overuse and potential deterioration of specific areas of the canvas, and it also provides a measurable framework for assessing system behavior: in order to enable further rigorous evaluation of the model, interaction data should at minimum be analyzed with respect to its deviation from a relatively even distribution of touches over time.

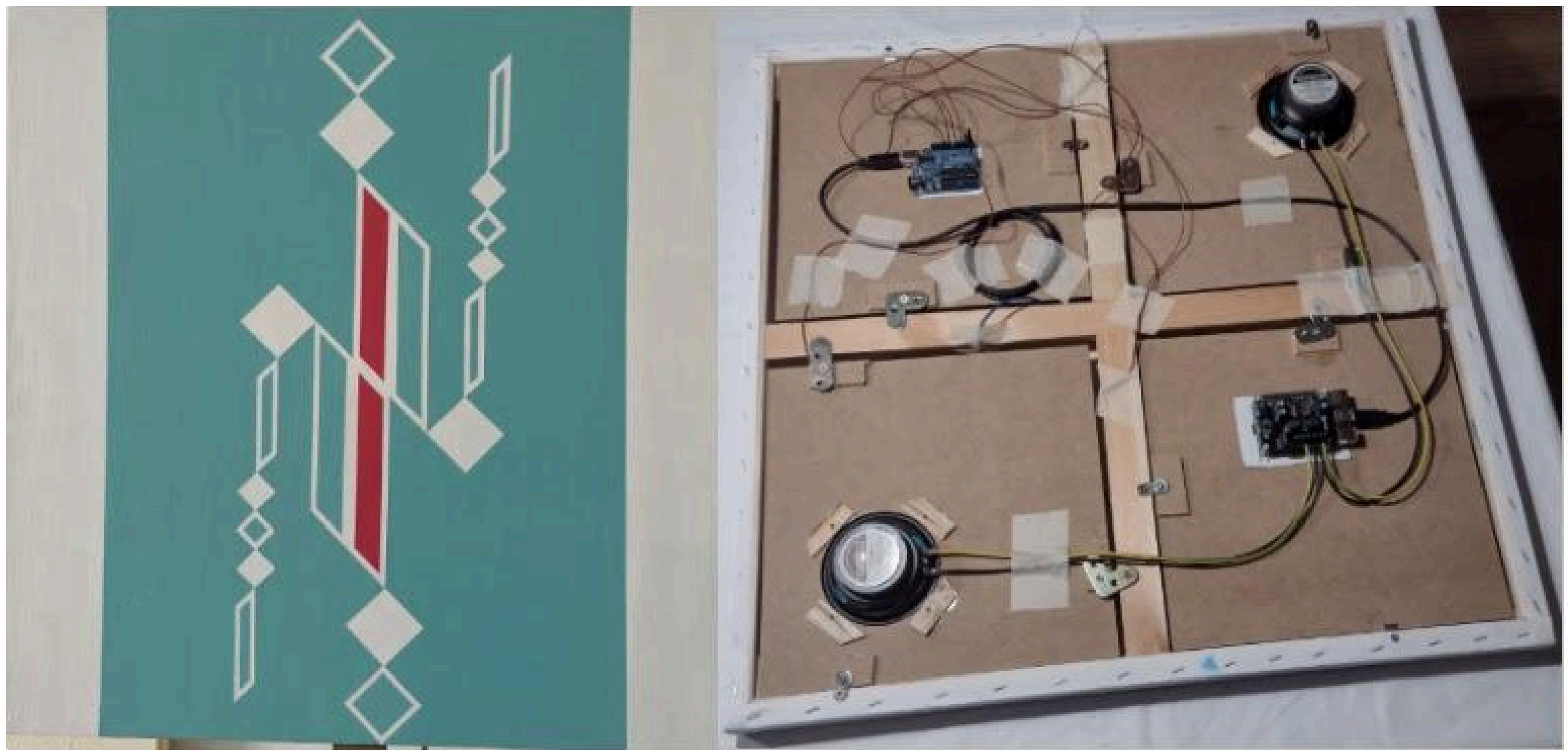

*Fig. 4: Front and Back views of "Echo of Lines"*

The networked architecture transforms individual performances into a collectively structured composition while preserving the autonomy of each Sounding Canvas. Rather than streaming audio between locations, the system transmits structured data describing gestures performed on each surface (e.g., position, velocity, and duration) via WebSockets. These data packets are interpreted locally by peer canvases, where they modulate internal parameters without replacing their own sonic material. Each Sounding Canvas therefore maintains a distinct sound world: participants hear the “voice” of their local canvas, not the audio output of a remote one.
In this framework, a touch in Rome does not reproduce its sound elsewhere; instead, it influences how another canvas generates its own response according to its specific aesthetic constraints. Interaction circulates as information rather than as sound, creating a distributed yet non-identical sonic ecology. LSTM networks integrate both local and network-derived inputs, enabling the system to adapt over time and to develop patterns shaped by collective activity. The resulting compositions emerge from the interplay between individual gestures, algorithmic mediation, and networked exchange, producing musical forms that extend beyond a single physical site while retaining distinct sonic identities.

Observing visitor behavior during recent Sounding Canvas exhibitions, one can identify the following possible scenario:
A visitor approaches the canvas tentatively, placing her palm against the upper-right quadrant. The capacitive sensors immediately detect the gesture, triggering the LSTM network to evaluate this initial contact against its learned interaction patterns. Rather than generating audio content, the algorithm selects from pre-clustered sound banks. Each sensor channel is permanently associated with one specific sound group, so when a given channel is activated, the algorithm can only select sounds from its corresponding cluster. The network recognizes this broad palm contact as an "exploratory gesture" and retrieves a corresponding ambient guitar sustain from Sensor 1's sound cluster. As the visitor experiments with different touch velocities and durations, the algorithm cross-references these gestural parameters with its embedded goals, selecting sounds that encourage continued exploration rather than overwhelming the interaction. After several minutes, the visitor develops rhythmic tapping patterns on the lower sensors. The LSTM identifies this gestural signature within its learned taxonomy and responds by accessing percussive sound clusters associated with rhythmic interaction modes. Crucially, the algorithm's temporal memory enables it to recognize when the visitor returns repeatedly to specific canvas regions, interpreting these patterns as preference indicators and weighting future sound selections accordingly.

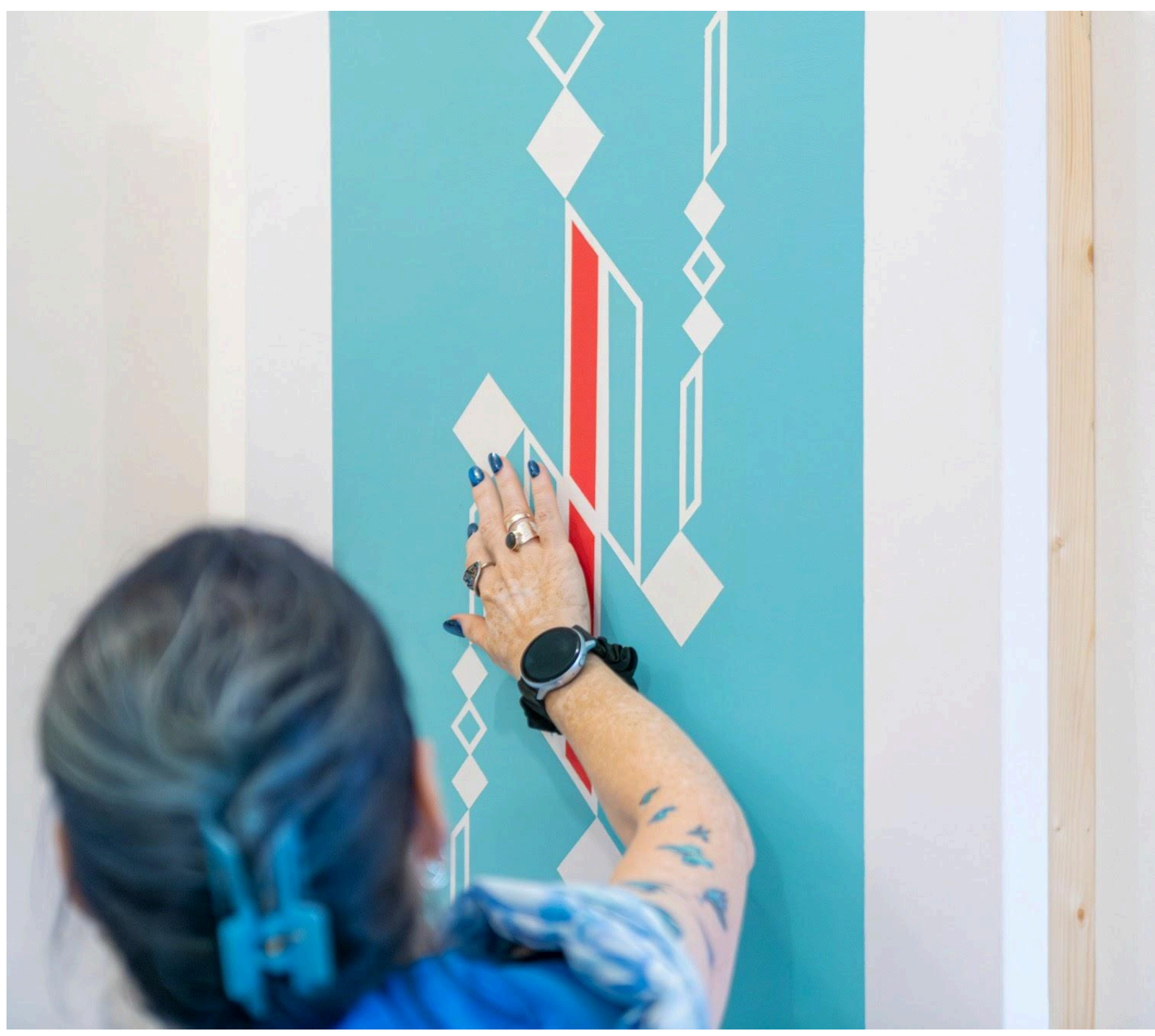

*Fig. 5: Sounding Canvas at Venice*

The distributed network architecture introduces an additional layer of complexity. When remote canvases transmit interaction data, the local algorithm incorporates this information into its selection logic. Visitors frequently report sensing "phantom collaborators", an uncanny awareness of remote presence mediated through the algorithm's contextual sound selection, where local gestures trigger responses that seem to acknowledge invisible partners across the networked system.

# 5. Algorithms as Artistic Material

The algorithmic architecture of the Sounding Canvas is structured around a distinction between off-line and on-line components. Off-line processes are those that operate prior to public interaction, shaping the latent space of possible correspondences between the visual and the sonic. In this category, Convolutional Neural Networks (CNNs) (LeCun et al. 1998) are employed to analyze the visual features of the painting, color, shape, and texture, and map them into a domain of sonic descriptors. This preparatory layer defines the structural vocabulary from which the artwork will subsequently draw.
On-line processes, by contrast, unfold in real time as the audience engages with the canvas. Here the system listens, adapts, and generates responses on the fly. Two families of models have been explored: Higher-Order Markov Models (HOMM) (Ching et al. 2004), which capture evolving probabilistic relationships between gestures and sounds, and Recurrent Neural Networks (RNNs) (Elman 1990), specifically Long Short-Term Memory (LSTM) (Hochreiter 1997) architectures, which differentiate between short-term fluctuations and longer-term patterns of interaction. Whereas the CNN defines the aesthetic substrate, the on-line models enact the performative layer, giving rise to an agent that evolves dynamically with its users.

This division between off-line and on-line components is not merely technical but aesthetic. The off-line algorithms encode the visual identity of each canvas into a set of musical descriptors, establishing a defined sonic framework and ensuring aesthetic consistency. The on-line LSTM models process live sensor input, differentiating short- and long-term interaction patterns to support adaptive behavior over time. Together, these layers structure the system as both preconfigured and responsive, enabling coherent yet evolving interactions without reducing the output to fixed mappings or arbitrary sound events.

The offline component of the Sounding Canvas employs a Convolutional Neural Network (CNN) (LeCun et al. 1998) to perform an unconventional but conceptually central task: mapping high-dimensional visual representations into the domain of sound descriptors.
To bridge the gap between these two different domains, visual features and audio descriptors, the system learns a linear transformation, represented by the matrix T. This matrix is trained using a paired dataset of image feature vectors and their corresponding sound feature vectors (see also Section 2 in which the processes for database creation are described). First, each image of the dataset, carefully preprocessed and normalized, is transformed into a 2048-dimensional feature vector that encodes the visual structure of the chosen video still from the dataset. Second, each associated audio segment is processed and characterized by a 10-dimensional vector of perceptually relevant features, including spectral centroid, rolloff, flux, bandwidth, flatness, zero-crossing rate, RMS energy, tempo, and simplified attack/decay times, extracted using the Librosa library (McFee et al. 2015). A dedicated python module in the CNN codebase uses the 2048-dimensional image vector $x$ , and the target 10-dimensional audio vector $y$ to calculate the transformation operator T. The matrix T is then optimized so that, when applied to $x$ via simple matrix multiplication, the result closely approximates $y$

$$y = T \cdot x$$

where $x$ is a 1×D tensor representing the feature vector of a single image (D = number of visual features extracted by the CNN), $y$ is the resulting predicted sound feature vector (1×O tensor, O = number of sound features) and T is the linear transformation matrix learned during training (of dimension OxD). Formally, the system minimizes the difference between $y$ and $T \cdot x$ across all training pairs using a mean squared error loss. During this optimization, each element of T is adjusted iteratively until the predicted audio vectors match the actual ones as accurately as possible. Once training is complete, T encodes the learned relationship between the visual and audio spaces: multiplying any new image feature vector by T produces a plausible corresponding audio feature vector.
Although more expressive nonlinear mappings such as multilayer perceptrons (MLPs) could in principle approximate more complex relationships between visual and audio features, a linear transformation was adopted deliberately. This choice enforces a direct and interpretable correspondence between the two domains, where each coefficient of (T) encodes the contribution of a specific visual feature to a given audio descriptor. In addition, the linear model provides greater robustness in limited-data conditions and avoids overfitting. This transformation is not merely a mathematical convenience; it embodies the artistic principle of duality, linking visual and sonic phenomena in a manner reminiscent of wave-particle complementarity. Once computed, the “T” matrix can be applied to new visual inputs, producing sound feature vectors that retain this duality.

Each canvas thus acquires a coherent identity: its visual surface and its generated soundscape are two manifestations of the same underlying structure. The CNN is used solely to extract meaningful, high-level visual features; the linear transformation then translates these features into a human-readable “score” providing the composer with a structured framework to generate sound, whether through instrumental performance, DAW manipulations, or algorithmic synthesis.

Image: calligraph.png

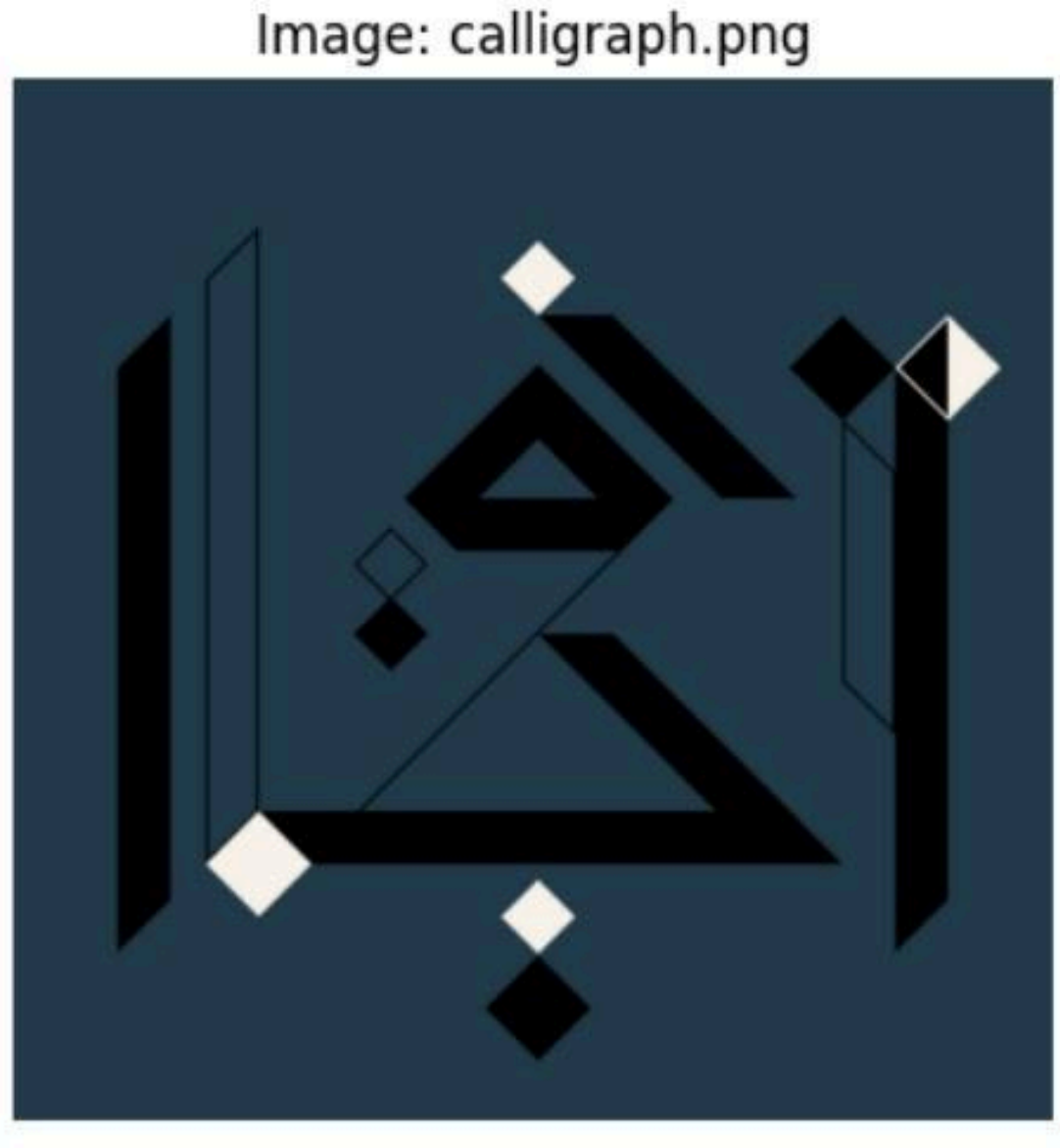

```
--- Predicted Sound Features ---
spectral_centroid_mean: 1150.4203
spectral_rolloff_85_mean: 2105.8812
spectral_flux_mean: 0.8942
spectral_bandwidth_mean: 1420.3310
spectral_flatness_mean: 0.0451
zero_crossing_rate_mean: 0.1245
rms_db_mean: -12.4052
tempo_bpm: 138.5000
attack_time_s: 0.0052
decay_time_s: 0.1504
```

Image: score.png

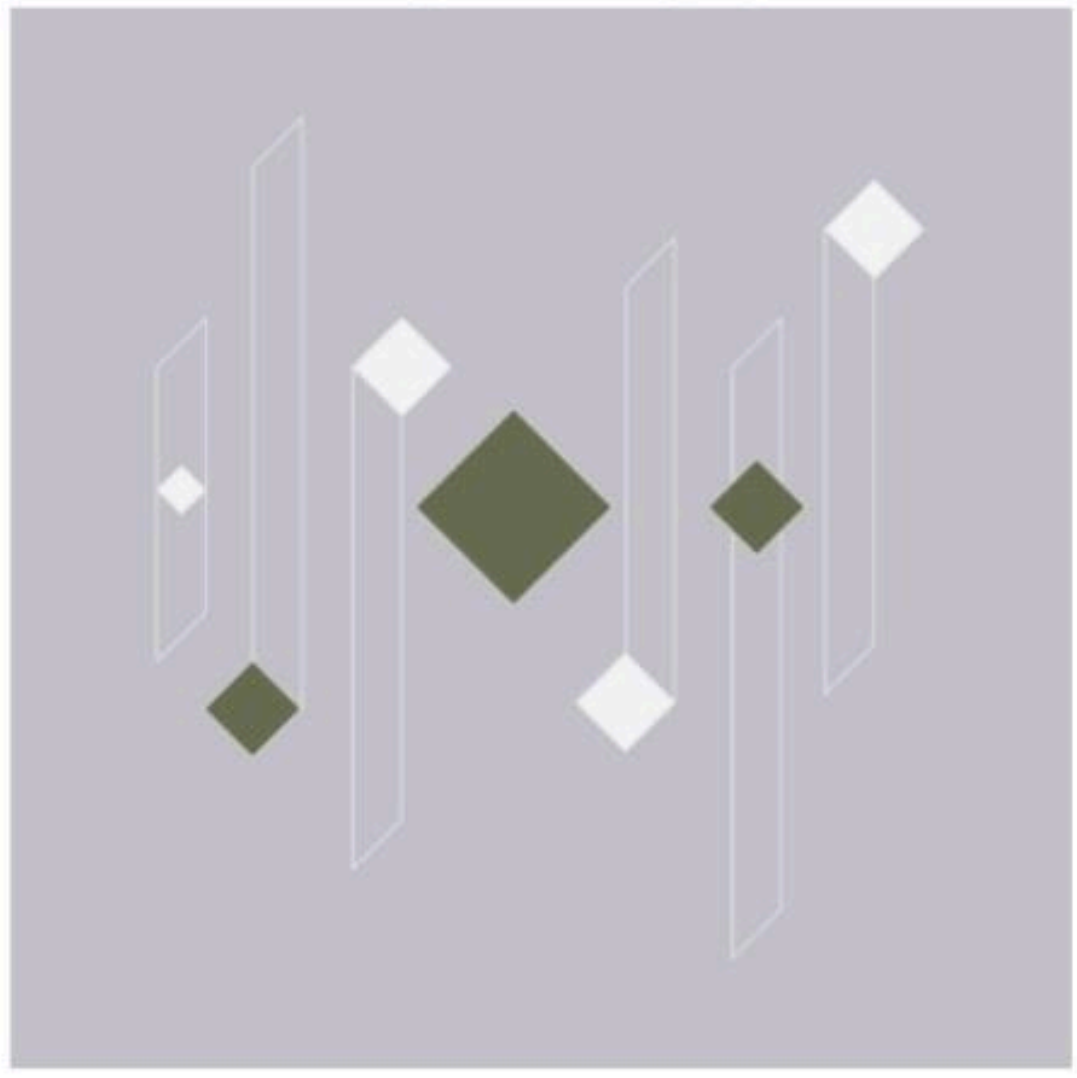

```
--- Predicted Sound Features ---
spectral_centroid_mean: 3420.1194
spectral_rolloff_85_mean: 5890.2231
spectral_flux_mean: 0.2104
spectral_bandwidth_mean: 2150.9923
spectral_flatness_mean: 0.0128
zero_crossing_rate_mean: 0.0412
rms_db_mean: -28.1140
tempo_bpm: 82.0000
attack_time_s: 0.1250
decay_time_s: 2.4501
```

*Fig. 6: Predicted descriptors by the CNN (the “score”) for the input canvas image*

While the CNN and “T” matrix operate offline, their output informs the real-time behavior of the online system: each Sounding Canvas interprets gestures and interactions through a sound vocabulary already grounded in the unique visual signature of its painting. In this sense, the transformation from image to sound is both a technical and an aesthetic act, unifying the visual and auditory realms and endowing each canvas with a distinct voice and look. In Figure 6 are shown the calculated descriptors (the “scores”) of two paintings.

The initial on-line component of the Sounding Canvas employed a High-Order Markov Model (HOMM) to manage real-time gesture-to-sound mapping. This system maintains a history of recent touch events up to a configurable order (default 8) and dynamically learns transition probabilities between sequences of interactions. Each new local event, representing a touch on a sensor channel, is encoded as a tuple and appended to the history:

$$\text{context} = (e_{t-n}, \ldots, e_{t-1})$$

The HOMM predicts the next sound using weighted transition counts:

$$P(s \mid \text{context}) = \frac{\text{count}(\text{context} \to s)}{\sum_{s'} \text{count}(\text{context} \to s')}$$

where counts decay over time

$$\text{count} \leftarrow \text{count} \cdot \text{decay}$$

to prioritize recent gestures. If a context has no recorded transitions, the model backs off to shorter contexts or a uniform distribution. After selecting a sound, the model updates its counts, reinforcing frequently traversed sequences while remaining adaptive. Remote events from other canvases are appended to the same history, enabling a networked, distributed dialogue between installations.

This HOMM approach allows each canvas to develop a contextual memory of interactions, producing temporally coherent yet flexible sonic responses. However, when interacting with a single user over extended periods, HOMM can tend toward self-reinforcement: without exposure to new contexts, the model may repeatedly favor the same sequences, reducing variability and limiting the sense of agency. To address this limitation, we explored the use of Recurrent Neural Networks (RNNs) (Elman 1990), specifically Long Short-Term Memory (LSTM) (Hochreiter 1997) models, as an alternative on-line component.

The LSTM-Based Event Manager serves as the on-line component with a specific goal: to replace HOMM when a single user interacts with a canvas over extended periods. Rather than reinforcing fixed transition tables, this component learns temporal patterns of dialogue between machine sounds and human touches, then uses those patterns to steer future interaction by nudging the user toward underused regions of the canvas.

For training data and supervision, interaction logs of past exhibitions are segmented into sessions, with each session consisting of a time-ordered list of events. Each event contains five key elements: the sound ID $s_i$ played by the system just prior to the user's touch (categorical, where S equals channels times sounds per channel), the inter-onset interval $\Delta t_i$ since the previous event, the touch duration $\tau_i$, the average gesture speed $v_i$, and the channel $c_i$ touched by the user at event i, which serves as the label. From each prefix $H_i$ containing the sequence of tuples from the first event through event i, the system creates one supervised example with target $c_i$. This approach trains the model to estimate

$$p_\theta(c_i | H_i)$$

for $c_i$ in the range from 1 to C, matching deployment conditions where the system hypothesizes the machine's next sound, appends it to the history, and asks the model which channel the human is likely to touch next. To support minibatching, histories are left-padded to a fixed length L (such as L=150), using zero IDs and zeros for continuous features, and because categorical and continuous streams are concatenated, masking is not applied to keep the pipeline simple and robust.

The architecture consists of four main components. First, inputs are processed through two streams: the categorical sequence $s_{1:L}$ is embedded via a trainable lookup $E \in \mathbb{R}^{S \times d}$ such that $e_t = E[s_t] \in \mathbb{R}^d$, while the continuous sequence $x_t = [\Delta t_t, \tau_t, v_t] \in \mathbb{R}^3$ is standardized if desired. Second, fusion occurs by concatenating per-time-step:

$$z_t = [e_t; x_t] \in \mathbb{R}^{d+3}$$

Third, a recurrent encoder using a single LSTM layer processes $z_{1:L}$ to produce a final hidden state $h_L \in \mathbb{R}^r$ that summarizes both short-term recent gestures and long-term session habits through

$$h_L = \mathrm{LSTM}_\theta(z_{1:L})$$

Fourth, a classifier employs a softmax layer to map $h_L$ to a categorical distribution over channels:

$$p_\theta(c|H) = \mathrm{softmax}(Wh_L + b)$$

where $W \in \mathbb{R}^{C \times r}$.

For loss and optimization, the system minimizes the negative log-likelihood (cross-entropy) over all examples:

$$\mathcal{L}(\theta) = -\sum_i \log p_\theta(c_i|H_i)$$ ,

using Adam optimization. With modest datasets, an embedding size d≈32 and hidden size r≈64 provide a good capacity-to-overfitting trade-off on embedded hardware.

```
{
  "metadata": {
    "description": "Training dataset for RNN",
    "created": "2025-07-12",
    "num_samples": 1234,
    "order": 8
  },
  "sequences": [
    {
      "canvas_id": "Echoes",
      "events": [
        {"type": "local", "channel_id": 1, "active_since": 0.23, "sound_id": 5, "timestamp": 0.53},
        {"type": "remote", "canvas_id": "rhythm", "channel_id": 2, "sound_id": 3, "timestamp": 1.12},
        ...
      ]
    },
    {
      "canvas_id": "script",
      "events": [
        {"type": "local", "channel_id": 3, "active_since": 0.41, "sound_id": 1, "timestamp": 0.88},
        ...
      ]
    }
  ]
}
```

*Fig. 7: structure of real-time logging data collected during exhibitions and used for training the LSTM model*

These architectural choices are motivated by several factors. LSTMs are preferred over GRU or Transformers because they are data-efficient, with explicit gating for long versus short-term dependencies, making them well-suited to small, heterogeneous logs and real-time inference on Pi-class devices, while Transformers would add latency, parameter count, and tuning overhead with little gain in this regime. The dual-stream input design captures both categorical sound ID information representing discrete machine utterances (bank/region, timbre family) and continuous timing/gesture features capturing rhythm and motor intent $(\Delta t, \tau, v)$, allowing their fusion to enable the model to learn patterns such as how short, fast touches after brighter sounds tend to migrate to specific channels. Left-padding without masks is employed because continuous features are also padded with zeros, and turning on embedding masks would require parallel masking for the numeric stream or custom layers, while empirically, the LSTM learns to ignore the leading zeros once sequences are sufficiently long, avoiding extra complexity and runtime cost.

At run time, every user touch on channel $c_t$ triggers a sequence of four key operations. First, the system performs a state update by computing the time since the last event ($\Delta t_t$) and appending the tuple $(s_t, \Delta t_t, \tau_t, v_t)$ to the interaction history, with the history being reset if the idle gap exceeds a specified threshold to avoid stale context. Second, the system defines a goal by maintaining per-channel counts and selecting a goal channel g with the smallest count to balance usage across the interface, implementing wear-leveling and spatial variety while using caps to prevent unbounded growth. Third, the system evaluates machine candidates by constraining potential sounds to the bank of the just-touched channel, preserving spatial coherence and the artwork's local voice. For each candidate sound a in that bank, the system forms a hypothetical next history

$$H_t \cup \{(a, \Delta t_t, \tau_t, v_t)\},$$

queries the model to obtain

$$p_\theta(c_{t+1}|H_t \cup \{a\}),$$

and records the probability of the goal channel $p_\theta(c_{t+1} = g|H_t \cup \{a\})$.

Fourth, the system selects with guidance and diversity by choosing

$$a^* = \arg\max_a p_\theta(c_{t+1} = g|H_t \cup \{a\}),$$

while maintaining a shortlist of all candidates that already predict g as most likely and sampling among them to preserve variety and avoid a single deterministic attractor.

In essence, the model estimates how the next human action will respond to each possible machine utterance, and the policy selects a sound that best advances the artistic goal of balanced exploration while maintaining stylistic continuity through the same spatial bank.

Regarding practicalities and hyperparameters, the vocabulary size is

$$S = C \times K + 1,$$

where C represents the number of channels and K represents sounds per channel, with the extra ID 0 reserved for padding. The sequence length L is set to 150 to cover typical sessions without bloating state, as longer sequences showed diminishing returns versus latency. Regularization through dropout and recurrent dropout can be enabled if logs grow, though initial deployments favored a lean model to keep inference responsive. Data balance considerations arise because users favor certain regions, creating class-imbalanced logs, which the goal-selection policy counteracts at inference time, with optional class-weighting available during training.

This approach proves artistically useful in several ways. Rather than being merely reactive like HOMM, which mirrors the recent past, the LSTM anticipates the user's next move conditioned on nuanced temporal cues, enabling the system to lead gently rather than only follow. The system demonstrates gesture sensitivity by treating timing and duration as first-class citizens, so the machine's response is shaped by how the user touches, not just where he/she touches. Additionally, the system maintains sustained variety without requiring external novelty, as even in single-user scenarios, the combined policy and model avoid self-reinforcing loops, preserving a sense of agency and dialogue over time. The approach does have limitations and suggests future extensions. Future work could incorporate reinforcement learning to optimize for multi-objective rewards encompassing variety, balance, and user engagement. Furthermore, adding confidence-aware strategies such as entropy-based exploration could adapt randomness to model certainty, further refining the balance between guidance and surprise.

The contrast between HOMM and LSTM approaches highlights how algorithmic choice directly affects the perception of machine agency. While HOMM emphasizes probabilistic reaction, LSTM introduces a sense of proactive authorship, allowing the artwork to participate in a shared creative process. When canvases are networked, each node contributes to a distributed, telepresent dialogue: the LSTM-based agent can propagate learned behaviors, fostering coherent yet evolving interaction patterns across geographically distant participants. In this way, the system exemplifies the role of the machine as co-creator, whose behavior is shaped both by accumulated experience and by the relational context of human engagement.

## 6. Discussion

The evolution of algorithmic sound art reveals a progression from autonomous generation toward increasingly sophisticated forms of human-machine collaboration. While pioneers like Lejaren Hiller established algorithms as pre-compositional tools that operated independently of real-time human input, contemporary practice has moved toward responsive systems that engage directly with user participation during live performance. Sounding Canvas represents a critical extension of this trajectory, transforming algorithms from reactive interfaces into proactive social mediators that enable distributed artistic collaboration across geographical boundaries. Unlike traditional algorithmic composition that emphasizes direct sound synthesis within isolated computational environments, this project pioneers an algorithmic curation approach where LSTM networks adaptively select from semantically organized sound banks based on learned user patterns and networked interaction data. The system's distributed architecture fundamentally reframes algorithmic authorship, moving beyond individual human-machine dialogues toward collective compositional processes where multiple adaptive algorithms coordinate to facilitate cross-continental artistic encounters. This paradigmatic shift raises fundamental questions about the nature of creative agency in networked artistic systems, the social implications of embedding machine learning into collaborative art practices, and the aesthetic potential of algorithms that function not merely as generative tools but as active participants in distributed creative communities that span physical and digital realms.

The Sounding Canvas project situates itself within the evolution of algorithmic sound art, exploring how learning-based embeddings, probabilistic modeling, and multimodal mapping can extend human-machine interaction. Whereas early generative works of the 1960s relied on explicit rules to transform visual or conceptual inputs into sound, our system implements an off-line Convolutional Neural Network (CNN) to map visual features of the canvas to audio descriptors. This replaces hand-coded rules with a learned, organic form of synesthesia, where the system discerns patterns and creates non-linear correspondences between sensory domains. In the on-line environment, the system employs a probabilistic Long Short-Term Memory (LSTM) model to capture patterns in user interactions, anticipating likely gestures and guiding sound generation towards a specific operational goal: balanced engagement across sensor channels. This departs from merely reactive systems, giving the algorithm intentionality and creating a nuanced dialogue with an adaptive agent. The resulting probabilistic landscape involves calculated system "choices" that subtly guide the user, forming a dynamic feedback loop. In doing so, Sounding Canvas provides a concrete instance of algorithms functioning as mediators in the creative process, addressing questions of algorithmic agency, anticipatory behavior, and the interplay between structured constraints and user freedom.

During exhibitions, users displayed diverse interaction patterns, ranging from localized exploration to broader gestures. Sensor activation logs show variability in gesture repetition and distribution over time, suggesting that interaction dynamics may influence system behavior. In networked sessions, concurrent interactions across canvases produced variations in the sonic output, although no formal analysis has been conducted to quantify these effects.

These preliminary observations, derived from system logs, motivate future work aimed at evaluating the relationship between user interaction patterns and the system's adaptive responses. At present, no definitive claims are made regarding the effectiveness or stability of the mapping, and a dedicated study will be required to assess these aspects rigorously.

In the Sounding Canvas, the algorithm serves as the central processing component, mapping sensor data (touch, gesture, movement) to sound in real time. Users engage with the system through tactile and auditory interaction. When multiple canvases are networked, interactions on one device can influence others, introducing a distributed dimension to the experience. Variations in the sonic output may reflect both local input and activity on connected canvases.

This configuration defines the Sounding Canvas as an interactive system in which human input and algorithmic processes are coupled. The resulting behavior can be interpreted as a co-evolving process, where local and remote interactions contribute to shaping the system over time, highlighting the relational aspects of networked, interactive artworks.

# 7. Conclusion

The Sounding Canvas integrates algorithms as structural components of the artwork rather than auxiliary tools. The off-line CNN embeddings establish a coherent mapping between visual features and sonic descriptors, defining each canvas's aesthetic vocabulary. In parallel, on-line processes, whether HOMM or LSTM-based models, govern the real-time evolution of interaction, interpreting gestures and adapting system behavior. Together, these layers couple sensing and computation, enabling each canvas to respond in a context-sensitive manner while maintaining a stable sonic identity. In networked configurations, this architecture extends interaction across participants, with algorithms mediating rather than replacing human contribution. The project demonstrates that algorithms can move beyond instrumental roles, functioning as central, perceptible agents that shape aesthetic outcomes, mediate social interaction, and define the identity and behavior of the artwork itself. This repositioning challenges conventional notions of authorship, interactivity, and agency in algorithmic art, emphasizing the creative and conceptual significance of embedding computation directly into the artwork's core.

Future work for the Sounding Canvas focuses on consolidating and extending its algorithmic foundations, particularly regarding interaction modeling and user adaptation. While the system currently employs two distinct event management approaches, Higher-Order Markov Models (HOMM) for multi-user exhibition scenarios and LSTM-based Recurrent Neural Networks (RNNs) for single-user interactions, these roles remain largely conceptual. A systematic evaluation is necessary to rigorously assess the hypothesis that HOMM better supports distributed, multi-user engagement, whereas RNNs excel at personalizing interactions to a single user's behavior. Such studies would provide empirical grounding for these design choices and inform potential refinements in the deployment strategy across different exhibition contexts.
Another important direction concerns personalized, user-specific adaptation. Building on the LSTM framework, future iterations could integrate richer user vectors capturing traits such as behavioral patterns, interaction style, and aesthetic preferences. These models would allow the canvas to respond not only to immediate touch gestures but also to the individual's history of engagement, subtly adapting sonic responses to enhance immersion and perceived agency. Complementing the current gesture-speed detection, additional modalities, such as touch pressure or contextual environmental data, could further refine the system's responsiveness, creating a nuanced feedback loop tailored to each user while preserving the emergent, semi-autonomous behavior of the canvas.
Longitudinal studies will play a central role in validating these extensions. By observing repeated interactions over weeks or months, researchers can investigate how user engagement evolves, how the system adapts to persistent patterns, and how users' perception of the canvas as a co-creative interlocutor develops over time. Such studies would also allow testing the potential for integrating long-term personalization while maintaining an aesthetically coherent interaction across sessions.
These developments could inform interdisciplinary applications beyond the gallery, including wellbeing practices, education, and collaborative creative tools. By embedding algorithms as active agents within the artwork itself, the Sounding Canvas fosters not only individual exploration and customization but also a sense of empathy toward the system and, when networked, toward other remote users. Future work in this direction would continue to position algorithms as central to both the technical operation and the aesthetic experience of the artwork, reinforcing the role of the canvas as a socially and sensorially responsive interlocutor.

The Sounding Canvas posits algorithms as integral artistic mediators, not art generators. Art isn't computationally produced; it emerges from the interconnectedness of beings, materials, actions, and experiences. Artistic creation is the recognition of resonant forms. Artists don't invent but perceive and clarify these emergent forms.

Computational processes, like CNN embeddings and recurrent networks, extend human perception and cognition. The Sounding Canvas uses algorithms as a lens, revealing latent structures in visual, sonic, and gestural interactions, enabling coherent experiences. Here computation amplifies perception and supports form recognition, becoming inseparable from the artwork's aesthetic structure.

This perspective explores truth, individuality, and consciousness. Self-experience arises from relationships with the world. Engaging with the Sounding Canvas highlights these relations: gestures, sounds, and networked interactions reflect both self and interconnectedness, fostering awareness of interdependence.

While free will and consciousness remain open questions, the Sounding Canvas offers a starting point: attending to relations, patterns, and interactions deepens our understanding of how phenomena coalesce into meaning. The artwork exemplifies a convergence of technical and aesthetic choices, where algorithms actively reveal hidden structures of perception, experience, and existence, aligning computation with the pursuit of insight and human understanding.

The project GitHub repository is available at:
https://github.com/luciamarock/SoundingCanvas

## References

Atkinson, R. C. and Shiffrin, R. M. 1968. Human Memory: A Proposed System and its Control Processes. In K. W. Spence and J. T. Spence (eds.) The Psychology of Learning and Motivation, Vol. 2. New York: Academic Press, 89–195.

Baddeley, A. D. and Hitch, G. 1974. Working Memory. In G. A. Bower (ed.) The Psychology of Learning and Motivation, Vol. 8. New York: Academic Press, 47–89.

Bentley, P. J. (ed.) 1999. Evolutionary Design by Computers. San Francisco, CA: Morgan Kaufmann Publishers.

Beuys, J. 1971. Energy Plan for the Western Man. Munich: Schirmer/Mosel.

Beuys, J. 1973. I Am Searching for Field Character. Düsseldorf: Kunstmuseum Düsseldorf.

Biles, J. A. 1994. GenJam: A Genetic Algorithm for Generating Jazz Solos. In Proceedings of the International Computer Music Conference. San Francisco, CA: International Computer Music Association, 131–137.

Blazey, R. 2017. Kalimbo: an Extended Thumb Piano and Minimal Control Interface. In Proceedings of the International Conference on New Interfaces for Musical Expression. Copenhagen: Aalborg University Copenhagen, 501–502.

Boulez, P. 1952. Schoenberg is Dead. Score 6 (May): 18–22.

Boulez, P. 1960s–70s. Notes on the Composition of Music. Various essays collected. Paris: Editions du Seuil.

Briot, A., Honnet, C. and Strohmeier, P. 2020. Stymphalian Birds – Exploring the Aesthetics of A Hybrid Textile. In Companion Publication of the 2020 Designing Interactive Systems Conference (DIS '20). New York, NY: ACM. DOI: https://doi.org/10.1145/3393914.3395840.

Bustos, F. and Pinto, H. S. 2022. Composing Music Inspired by Sculpture: A Cross-Domain Mapping and Genetic Algorithm Approach. Entropy 24(4): 468.

Cage, J. 1957. Experimental Music: Composition as Research. Tempo 43 (September): 7–17.

Cage, J. 1961. Silence: Lectures and Writings. Middletown, CT: Wesleyan University Press.

Carnovalini, F. and Rodà, A. 2020. Computational Creativity and Music Generation Systems: An Introduction to the State of the Art. Frontiers in Artificial Intelligence 3: 14.

Ching, W. K., Fung, E. S. and Ng, M. K. 2004. Higher-Order Markov Chain Models for Categorical Data Sequences. Naval Research Logistics 51(4): 557–574.

Crick, F. 1994. The Astonishing Hypothesis: The Scientific Search for the Soul. New York: Scribner.

Dannenberg, R. B. 2021. Communication for Real-Time Music Systems: An Overview of O2. Computer Music Journal 45(4): 7–19.

Elman, J. L. 1990. Finding Structure in Time. Cognitive Science 14(2): 179–211.

Fernandez, G. B. and Larson, K. 2021. Tune Field. In Proceedings of the International Conference on New Interfaces for Musical Expression. Online. DOI: 10.21428/92fbeb44.2305755b.

Fiebrink, R. and Cook, P. R. 2010. The Wekinator: A System for Real-time, Interactive Machine Learning in Music. In Proceedings of the International Society for Music Information Retrieval Conference, 413–418.

Fink, L., Fiehn, H. and Wald-Fuhrmann, M. 2024. The Role of Audiovisual Congruence in Aesthetic Appreciation of Contemporary Music and Visual Art. Scientific Reports 14: 20923.

Hochreiter, S. and Schmidhuber, J. 1997. Long Short-Term Memory. Neural Computation 9(8): 1735–1780.

Honigman, C., Hochenbaum, J. and Kapur, A. 2014. Techniques in Swept Frequency Capacitive Sensing: An Open Source Approach. In Proceedings of the International Conference on New Interfaces for Musical Expression. London: Goldsmiths, University of London, 74–77.

Koblin, A. and Echelman, J. 2014. Unnumbered Sparks. Installation premiered at TED Conference, Vancouver.

LeCun, Y., Bottou, L., Bengio, Y. and Haffner, P. 1998. Gradient-based Learning Applied to Document Recognition. Proceedings of the IEEE 86(11): 2278–2324.

Libet, B., Gleason, C. A., Wright, E. W. and Pearl, D. K. 1983. Time of Conscious Intention to Act in Relation to Onset of Cerebral Activity (Readiness-potential). Brain 106(3): 623–642. https://doi.org/10.1093/brain/106.3.623

Lozano-Hemmer, R. 2009. Corpus: Interactive Installations. Linz: Ars Electronica Center.

Lupone, M. and Galizia, L. 2006. Volumi Adattivi [interactive sound and sculptural installation]. Rome: Goethe Institut, International Biennale "Arte Scienza".

McFee, B., Raffel, C., Liang, D., Ellis, D. P. W., McVicar, M., Battenberg, E. and Nieto, O. 2015. librosa: Audio and Music Signal Analysis in Python. In Proceedings of the 14th Python in Science Conference, 18–25.

Murray-Browne, T., Mainstone, D., Bryan-Kinns, N. and Plumbley, M. D. 2018. Cave of Sounds. Interactive installation. London: Barbican Centre.

Neef, N., Zabel, S., Papoli, M. and Otto, S. 2024. Drawing the Full Picture on Diverging Findings: Adjusting the View on the Perception of Art Created by Artificial Intelligence. AI & Society 40: 2859–2879.

Raftery, A. E. 1985. A Model for High-Order Markov Chains. Journal of the Royal Statistical Society: Series B (Methodological) 47(3): 528–539.

Rutz, H. H. 2016. Agency and Algorithms. Journal of Science and Technology of the Arts 8(1): 5–19.

Rutz, H. H. 2017. Sound Processes: A System for Composing and Performing with Sound in Time. In Proceedings of the International Conference on New Interfaces for Musical Expression. Copenhagen: Aalborg University Copenhagen, 221–226.

Salnikov, V., Schaub, M. T. and Lambiotte, R. 2016. Using Higher-Order Markov Models to Reveal Flow-Based Communities in Networks. Scientific Reports 6: 23194.

Santos, L., Dionisio, M. and Campos, P. 2024. TapeStory: Exploring the Storytelling Potential of Interactive Tapestries. In Proceedings of the 16th Conference on Creativity & Cognition (C&C '24). New York, NY: ACM. DOI: https://doi.org/10.1145/3635636.3656211.

Santos, L., Olim, S., Campos, P. and Dionisio, M. 2025. Interactive Tapestry to Raise Marine Noise Pollution Awareness Among Teens. In Proceedings of the 24th Interaction Design and Children (IDC '25). New York, NY: ACM. DOI: https://doi.org/10.1145/3713043.3731516.

Todd, P. M. and Werner, G. M. 1999. Frankensteinian Methods for Evolutionary Music Composition. In N. Griffith and P. M. Todd (eds.) Musical Networks: Parallel Distributed Perception and Performance. Cambridge, MA: MIT Press, 313–339.

Visi, F. G. and Tanaka, A. 2020. Interactive Machine Learning of Musical Gesture. arXiv preprint arXiv:2011.13487.

Xenakis, I. 1959. Analogique B. Paris: Boosey & Hawkes.

Xenakis, I. 1971. Formalized Music: Thought and Mathematics in Composition. Bloomington and London: Indiana University Press.

## Web Sources

ALMAT 2020 (Algorithms that Matter). Online symposium on algorithmic agency in artistic practice. https://www.researchcatalogue.net/view/921059/921060 (accessed September 13th 2025).

Buechley, L. Tinkering Tinkerer. Large-scale interactive painting. San Francisco: Exploratorium. https://www.exploratorium.edu/tinkering/tinkerers/leah-buechley (accessed April 20th 2026).

Ciamarone, L. 2024. Sounding Canvas source code. GitHub. https://github.com/luciamarock/SoundingCanvas (accessed September 13th 2025).

Reuter, J. 2005. SoundPaint – Painting Music. In Proceedings of the 3rd International Linux Audio Conference (LAC2005). Karlsruhe: ZKM.

Simon Doury et al. 2021. Paint With Music. Google Magenta. https://magenta.withgoogle.com/paint-with-music (accessed September 13th 2025).

Soundwall. 2015. Chris Alker, 'Soundwall': The Connected Canvas. Insomniac Magazine. https://www.insomniac.com/magazine/soundwall-the-connected-canvas/ (accessed September 13th 2025).

Voto, D. 2025. Multisensory Interactive Installation. Noema. https://noemalab.eu/ideas/essay/multisensory-interactive-installation/ (accessed September 13th 2025).

Zhang, A. 2019. 'Conduct San Jose' Interactive Music Mural. The Tech Museum blog. https://medium.com/@TheTech/the-music-mural-c785586f1b2a (accessed September 13th 2025).